\documentclass{article}

\usepackage{authblk}

\usepackage[english]{babel}

\usepackage[letterpaper,top=2cm,bottom=2cm,left=3cm,right=3cm,marginparwidth=1.75cm]{geometry}

\usepackage{amsmath}
\usepackage{graphicx}
\usepackage{tikz}
\usetikzlibrary{arrows.meta,calc,positioning}
\usepackage[colorlinks=true, allcolors=blue]{hyperref}

\author{Matthew W. Cotton\thanks{mwc38@cam.ac.uk}}
\author{David Klenerman}
\author{Georg Meisl\thanks{gm373@cam.ac.uk}}
\affil{\textit{Yusuf Hamied Department of Chemistry, University of Cambridge, UK and
UK Dementia Research Institute at University of Cambridge, UK}}

\title{In Vivo Length Distributions as Mechanistic Fingerprints of Pathological Protein Aggregation}

\date{September 22, 2026}

\begin{document}
\maketitle

\begin{abstract}
Modern imaging techniques can resolve individual pathological protein aggregates in postmortem human samples, providing detailed measurements of aggregate size distributions that are inaccessible with conventional bulk approaches. These distributions represent mechanistic fingerprints of the microscopic processes that generated the observed pathology, but extracting this mechanistic information requires a quantitative theoretical framework. Here, we develop the mathematical tools needed to interpret aggregate length distributions in living systems, where aggregate growth competes with active removal. We show that, across a class of models, the length distribution of sufficiently large aggregates approaches a geometric decay. Crucially, the decay rate is determined by the balance between aggregate elongation and removal, providing a direct quantitative readout of these competing processes from a single time point measurement. This enables mechanistic comparisons between healthy and diseased human samples without requiring longitudinal measurements of aggregate dynamics. We further analyse how additional aggregation and removal processes modify the observed length distributions. Together, these results establish the mathematical foundations and tools to use aggregate length distributions as an experimentally accessible route for inferring microscopic aggregation dynamics directly from human tissue.

\end{abstract}

\section{Introduction}
Neurodegenerative diseases are a major cause of progressive loss of brain function and dementia. In these diseases, normally functional proteins come together to form large, fibrillar protein aggregates, often consisting of many thousands of protein monomers~\cite{Chiti2017, Knowles2014}. Although the proteins involved differ between diseases, the accumulation of misfolded protein aggregates is a defining pathological feature of many common neurodegenerative disorders, including Alzheimer’s disease, Parkinson’s disease and related proteinopathies. Understanding the processes that govern the formation, persistence and removal of these aggregates is therefore central to understanding how pathology emerges and progresses.

Protein aggregation has been studied extensively in vitro, where experiments can be initiated from well-defined protein concentrations and followed continuously over time~\cite{Cohen2012}. Combined with mathematical models of chemical kinetics, these measurements have produced a detailed understanding of the microscopic processes through which soluble proteins are converted into aggregates, including nucleation, elongation, fragmentation and aggregate-catalysed self-replication. By fitting kinetic models to time-resolved measurements, the relative contributions and rates of these processes can often be determined quantitatively~\cite{Meisl2016, Meisl2022func}.

Translating this understanding to human disease is considerably more difficult. The intracellular environment is not a closed reaction vessel: proteins are continually synthesised and degraded, while aggregates are acted on by active protein-quality-control and clearance pathways. The competition between aggregate formation and removal can fundamentally alter the behaviour of the system, allowing low levels of aggregation to persist in a stable state rather than inevitably progressing towards complete conversion of the available protein~\cite{Meisl2024}. More generally, it remains unclear which of the molecular processes identified under controlled in vitro conditions dominate aggregate accumulation in the human brain, and how these processes change between health and disease.

A second, fundamental difficulty is observational. The kinetics of aggregation in the human brain cannot normally be measured longitudinally at the molecular or cellular scale. Postmortem tissue provides detailed molecular information, but only at a single time point, while measurements in living individuals generally lack the spatial resolution required to resolve individual protein aggregates. The central challenge is therefore to solve an inverse problem, inferring the dynamics that generated a pathological state from a static measurement of the aggregates that remain.

Recent advances in single-molecule and super-resolution microscopy now provide an important opportunity to address this problem~\cite{Boeken2025}. These approaches can resolve large populations of individual aggregates directly in human tissue and tissue-derived samples, providing quantitative measurements not only of aggregate abundance but also of properties such as aggregate size. The resulting distributions contain substantially more information than a bulk measurement of aggregate concentration. In particular, as we show here, the relative abundance of aggregates of different lengths records the competition between the processes that alter aggregate size, including their growth, shrinkage, fragmentation and removal. Aggregate size distributions can therefore be viewed as a mechanistic fingerprint of the underlying aggregation dynamics.

Size distributions are also an attractive experimental observable. Measurements of individual aggregates can be obtained from extracted material without requiring an intact cellular or tissue environment, and the lengths of sufficiently stable aggregates can remain measurable across a range of sample-preparation procedures~\cite{Fertan2026PD}. Consequently, aggregate-length measurements can in some settings be obtained more readily and at substantially higher throughput than detailed information about aggregate ultrastructure, cellular identity or spatial localisation. This makes them a potentially powerful bridge between experimentally accessible measurements of human tissue and molecular models of aggregation.

In previous work, we demonstrated the potential of this approach using a minimal model in which aggregate growth competes with aggregate removal~\cite{English2026, Boeken2026}. Under simplifying assumptions, the model predicts a geometric aggregate-size distribution whose decay rate reports the relative balance between formation and removal. Applying this description to aggregate-size measurements from human brain samples showed that the distributions differ systematically between healthy and diseased tissue and that these differences can be used to infer changes in the underlying aggregation–removal balance~\cite{English2026}. This provided a proof of principle that mechanistic information can be recovered directly from aggregate-size statistics, even in the absence of a longitudinal measurement.

Here, we go beyond this deliberately simple description and develop a more complete mathematical framework for the general interpretation of experimental length distributions. We establish when particular distributions arise, how they depend on the microscopic processes acting on aggregates, and which mechanistic quantities can be identified robustly from finite experimental data.
We derive and analyse models for the evolution and steady-state distribution of aggregate lengths under a range of aggregation and removal mechanisms, and determine how experimentally observable features of these distributions relate to the underlying molecular rates. In doing so, we establish the conditions under which aggregate-length distributions can be used to distinguish mechanisms and quantify changes between biological states. Our aim is to turn an experimentally accessible quantity of postmortem human aggregates into a quantitative readout of the processes that generated them. More broadly, this provides a route for extracting information on dynamic processes from static human tissue and for connecting the extensive mechanistic understanding developed in vitro with the molecular processes operating in human neurodegenerative disease.

\section{Models of Aggregate Production and Removal}

\subsection{Generalised Master Equation and Moments}

We begin by writing down the general master equation that describes the evolution of the concentration of aggregates of size $i$ in a well mixed system, denoted by $f(i)$. Aggregates are modelled as linear chains of a monomeric protein and we denote the free monomer concentration (i.e $f(1)$) as $m$, as in earlier work~\cite{Knowles2009}. As detailed in prior work~\cite{Knowles2009, Cohen2011a,Cohen2011b, Cohen2011c, Meisl2017a}, several classes of processes lead to the interconversion of different aggregate sizes, including: a primary nucleation process (monomers come together to form an aggregate), a secondary nucleation process (monomers come together on the surface existing aggregate), elongation of aggregates (existing fibrils grow by monomers adding to their ends) and dissociation of aggregates (existing fibrils shrink by monomers dissociating from their ends). Additionally, we allow aggregates to be removed from the system as part of homeostasis within living systems, for example, passive transport or active removal mechanisms~\cite{Meisl2024, Thompson2021, Cotton2026JCP}. All together, the resulting chemical master equation that describes the system is
\begin{equation}
\begin{split}
    \frac{\text{d}f(i)}{\text{d}t} &= \delta_{i, n_C}k_n m^{n_C} + 2k_{+} m (f(i-1)-f(i))\\
    &- 2k_{\text{off}} (f(i)-f(i+1))  + \delta_{i, n_2}k_2 m^{n_2} \left( \sum_{j=n_c}^{\infty}jf(j)\right) \\
    &- k_{-} \left((i-1)f(i) - 2\sum_{j=i+1}^{\infty}f(j)\right) - c_i
\end{split}
\label{eq:master_eq}
\end{equation}
where the rate constants $k_n$, $k_2$, $k_{+}$, $k_{\text{off}}$ and $k_{-}$ correspond to the primary nucleation, secondary nucleation, polymerisation (elongation), depolymerisation and fragmentation processes respectively. The removal rate of aggregates of size $i$ is $c_i$ which may depend on a number of system parameters but will typically be first order in the aggregate concentration so that $c_i = r_i f(i)$ where $r_i$ is the rate of removal.

These dynamics produce a system of coupled equations that define the distribution of aggregates~\cite{Knowles2009}. We can define the moments of this distribution: the total mass of aggregates $M=\sum_i i f(i)$ and the total number of aggregates, $P=\sum_i f(i)$. Summing over equation \eqref{eq:master_eq} from $n_c$ to $\infty$ gives and simplifying as detailed in Cohen et al.~\cite{Cohen2011a} gives
\begin{align}
    \frac{\text{d}P}{\text{d}t} &= k_{-}M + k_{2}m^{n_2}M + k_{n}m^{n_{c}} - \sum_{i}^{\infty} r_i f(i).
\end{align}
and
\begin{align}
    \frac{\text{d}M}{\text{d}t} &= 2\left(k_{+}m - k_{\text{off}} \right) P - \sum_{i}^{\infty}i r_i f(i).
\end{align}
where we have assumed the rates of dissociation of aggregates of size $n_c$ 
and the contribution of nucleation to the aggregate mass are negligible (for details see Cohen et al.~\cite{Cohen2011a}). The final equation is to describe the monomer concentration $m(t)$. Since we focus on the case of protein aggregation in living systems, we keep this at a constant value $m(t)=m_0$ which will be maintained by cellular homeostasis~\cite{Thompson2021,Cotton2026JCP}.

\subsection{Signatures of Aggregate Removal}

In order to proceed we make two weak assumptions about the removal process. The first assumption is that there is a maximum rate at which aggregates can be removed from the system. We expect aggregate removal to saturate as aggregate burden increases. This is a relatively weak physical assumption, arising from limits on the abundance of the molecular machinery involved, the energy available to drive removal, and the rate at which the intermolecular bonds stabilising aggregates can be disrupted. The second assumption is that the rate of removal is independent of aggregate size. While this may not hold in all systems, it is a valid approximation when small variations in size are considered, and produces a closed set of moment equations. Scenarios with a size-dependent removal rate are discussed in previous work~\cite{Thompson2021, Cotton2026JCP}. An example of such removal that satisfies these assumptions is given by the Michaelis-Menten-like reaction kinetics. By considering a molecular clearance component that binds to aggregates~\cite{Cotton2026JCP}, the aggregate removal is independent of size such that
\begin{equation}
    r_i  = \frac{\lambda }{1+M/K_{M}} \equiv r
    \label{def_r}
\end{equation}
where $K_M$ determines the concentration of aggregates and we have defined $r$ as the size-independent removal rate.

In our earlier work~\cite{Cotton2026JCP}, we showed such saturating removal competes with aggregate formation to give rise to a characteristic bifurcation structure. At low monomer concentrations, the system has a stable low-aggregate steady state and an unstable steady state that defines the boundary beyond which aggregate formation overwhelms removal. As the monomer concentration increases, these two steady states approach one another and disappear through a saddle-node bifurcation, above which no stable low-aggregate state exists and the system undergoes runaway aggregation. This behaviour provides a natural explanation for several features observed in cellular aggregation systems, including stable low levels of aggregation, threshold-dependent seeding, and rapid transitions to a high-aggregate state. Cotton et al.~\cite{Cotton2026JCP} discusses how different assumptions affect the system, but shows that the emergence of two stable states is robust across different scenarios.

\begin{figure} [h]
    \centering
    \includegraphics[width=0.8\linewidth]{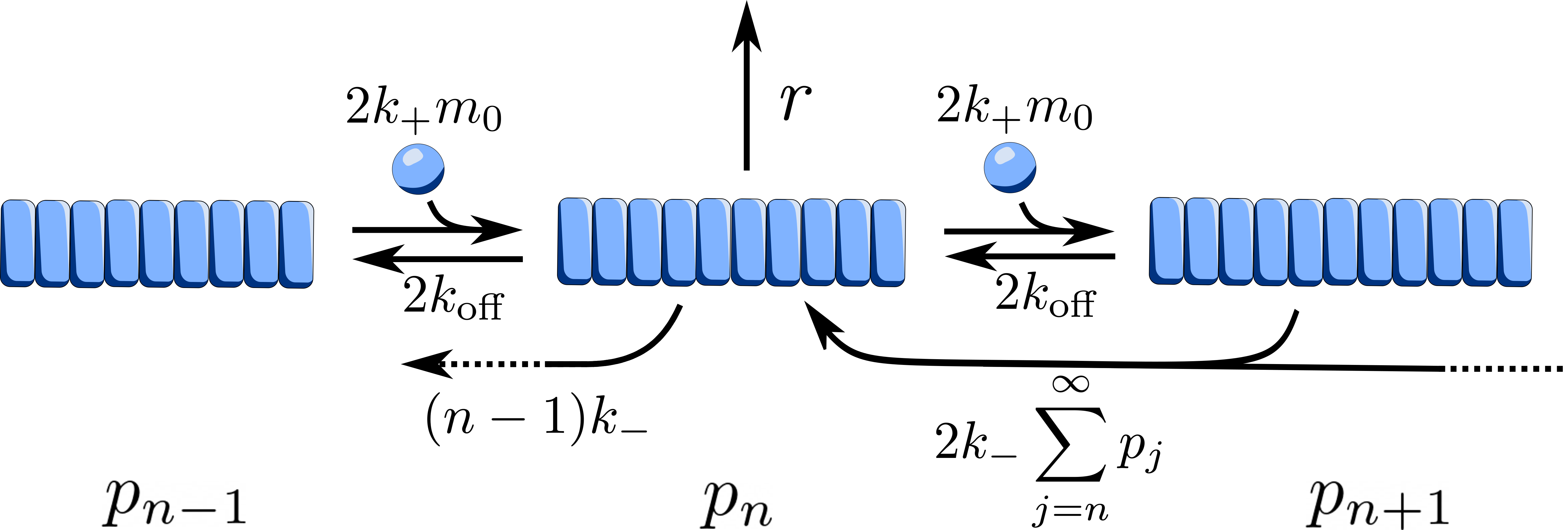}
    \caption{The population distribution of large aggregates are not governed by nucleation processes, but by processes that convert between aggregates of different sizes.}
    \label{fig:genrates}
\end{figure}

The two state model predicts a low aggregate load even when cells are in a healthy state. Since this state is stable, the aggregate length distribution will evolve until it relaxes to some constant distribution. In postmortem samples, we expect to see exactly this distribution, which mathematically corresponds to the right hand side of equation \eqref{eq:master_eq} vanishing. This means that the concentration flux into and out of the state of each size is balanced, see Figure \ref{fig:genrates}, however this is still a dynamic state as aggregates are being removed at each state and replenished by the growth of smaller aggregates. In systems where the aggregate mass is not constant, the normalised length distribution can still converge to some attracting distribution~\cite{Michaels2015}, and thus the length distributions can still provide insights into the distinct molecular processes even in these cases. In fact, by studying normalised length distributions rather than absolute concentrations, we are able to more reliably compare different experimental systems which might show different sensitivity or measure aggregates at different sample dilutions, as these effects do not alter aggregate sizes, only their absolute quantification.

\section{Size distributions in different regimes}

For the remainder of this work, we focus on description of the size-distribution of aggregate sizes that are well above the nucleation size. In an idealised system aggregates nucleate at size $n_{c}$ and $n_{2}$, however the real nucleation dynamics may cause aggregates of adjacent sizes to also be produced~\cite{Dear2018}. Practically, these nuclei will usually contain only on the order of 10 or fewer monomers, whereas mature aggregates contain hundreds to thousands of monomers. In the context of applying the findings of this work to experimental data, reliable data will likely be available only for aggregates above a certain size from super resolution microscopy (the technique usually requires the binding of multiple labelled antibodies to the same structure, and the minimum resolution is tens of nms). Furthermore, conformational changes in aggregate structure may occur as aggregates grow beyond their nucleation sizes, complicating the size distribution at small sizes. Therefore, we here consider the statistics of larger aggregates and the aggregation dynamics that govern them. At these sizes, measurements are more reliable and the size distribution changes predominantly by inter-conversion of aggregates of different sizes (via elongation, depolymerisation and fragmentation) or by complete removal of aggregates from the system. By limiting our discussion to the dynamics of large aggregates, the assumptions made about the uniformity of growth, fragmentation and removal are likely to be more accurate (for example, the kinetics of removal for an aggregate made of 100 monomers is likely to be very similar to an aggregate of 105 monomers, whereas the behaviour of a smaller aggregate of 5 monomers might differ significantly from one containing 10 aggregates). We now consider the resulting aggregate length distributions for different limiting cases below.

\subsection{Elongation and Removal only}

Firstly, we consider the regime of negligible depolymerisation ($k_{\text{off}}=0$), and negligible fragmentation ($k_{-}=0$). In this scenario aggregates nucleate, by either primary or secondary nucleation, and grow continually  via addition of monomeric protein, until removal mechanisms remove them from the system. As such, if we consider the long time solution of a steady state distribution, the system will reach a quasi-steady size distribution where the elongation process will be balanced by removal. In this limit equation \ref{eq:master_eq} for large $i$ becomes
\begin{equation}
    \frac{\text{d}f(i)}{\text{d}t} = 2k_{+} m_0 (f(i-1)-f(i)) - rf(i).
\end{equation}
Note that nucleation only produces fluxes into $i=n_c$ and $i=n_2$, so these terms do not contribute at large $i$. The stationary distribution of this system is a geometric distribution
\begin{equation}
    f(i) \propto \alpha^{i}
\end{equation}
with
\begin{equation}
    \alpha=\frac{1}{\frac{r}{2k_{+}m_0}+1}.
    \label{eq:alphadef}
\end{equation}

\begin{figure} [h]
    \centering
    \includegraphics[width=0.8\linewidth]{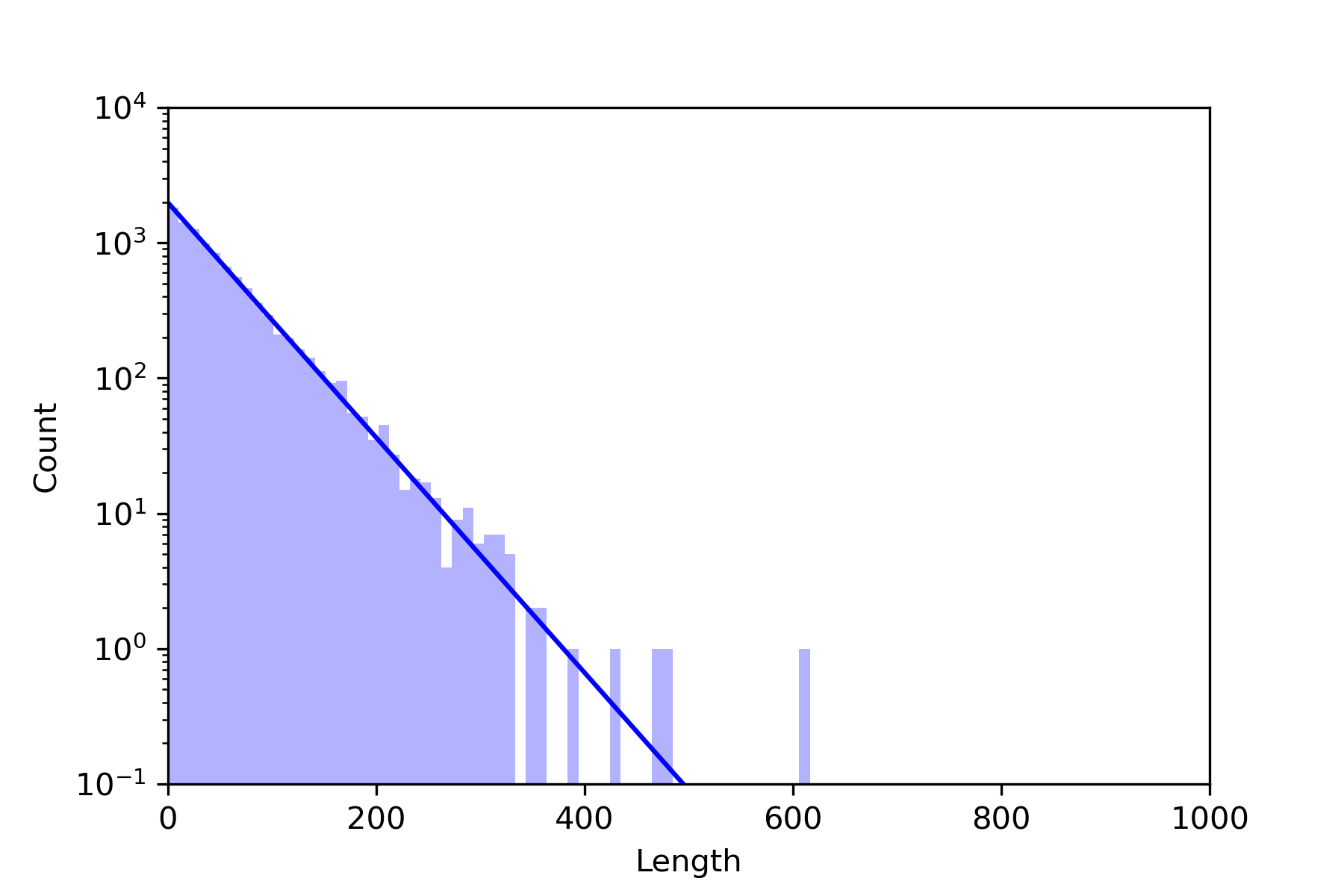}
    \caption{An artificially generated length distribution with $\alpha = 0.98$. On a logarithmic y-axis this geometric decay gives a straight line, thus this way of plotting provides a simple check that can be performed on experimental data.}
    \label{fig:example_len_dist}
\end{figure}

Changes to the overall nucleation rate will affect the overall aggregate mass but not change the normalised distribution. On a log-linear histogram showing the frequency of aggregates at every size, the geometric distribution will appear as a straight line, with slope $\log(\alpha)$, see Figure~\ref{fig:example_len_dist}.

The \textit{relative removal}, $\tilde{r}$, which we define as the ratio of the removal to elongation rate, can be calculated as $\tilde{r}\equiv r/(2k_{+}m_0)=\alpha^{-1}-1$ from equation~\eqref{eq:alphadef}. This quantity can then be used to compare the balance of elongation and removal mechanisms across different aggregate populations. 

\subsection{Depolymerisation}
The first extension we make to the minimal system is to include the depolymerisation process, the reverse of the elongation process. This process will become important when the monomer concentration approaches its solubility and there is little turnover of aggregated species by removal. The master equation for large $i$ now becomes

\begin{equation}
    \frac{\text{d}f(i)}{\text{d}t} =  2k_{+} m (f(i-1)-f(i))
    - 2k_{\text{off}} (f(i)-f(i+1))- rf(i)
\label{eq:pi_evolution_remove}
\end{equation}
We recover the following equation by setting the fluxes into and out of state $i$ to be equal:
\begin{equation}
    2k_{+} m_0 f(i-1) + 2k_{\text{off}} f(i+1) = (2k_{+} m_0 + 2k_{\text{off}} + r)f(i).
\end{equation}
This can again be solved by a geometric series ansatz, $f(i)\propto\alpha^i$, resulting in a quadratic equation in $\alpha$ solved by
\begin{equation}
    \alpha = \frac{r +2k_{\text{off}}+2k_{+} m_0}{4 k_{\text{off}}}\left(1 - \sqrt{1-\frac{16 k_{\text{off}} k_{+} m_0}{(r +2k_{\text{off}}+2k_{+} m_0)^2}}\right)
    \label{eq:alpha_depol}
\end{equation}
where we have only kept the negative root which gives the physical $\alpha < 1$ case. In systems where the length distribution is mainly determined by the aggregation and removal terms, we can expand this expression in the limit where the depolymerisation rate is slow compared to the other processes, i.e. $2k_{\text{off}} \ll r +2k_{+} m_0$. This gives
\begin{equation}
    \alpha =  \frac{2k_{+} m_0}{r + 2k_{+} m_0} - \frac{4 r k_{\text{off}} k_{+} m_0}{(r +2k_{+} m_0)^3} + \mathcal{O}\left(\left(\frac{2k_{\text{off}}}{r +2k_{+} m_0}\right)^2\right).
\end{equation}
In the limit of $k_{\text{off}} \to 0$ the second term vanishes and we recover the elongation and removal only case. As intuitively expected, in this regime the depolymerisation acts to reduce $\alpha$ and lower the concentration of larger aggregate species.

\subsection{Fragmentation}
Next we also consider the fragmentation of existing aggregates. The master equation for large $i$ is now
\begin{equation}
\begin{split}
    \frac{\text{d}f(i)}{\text{d}t} &=  2k_{+} m (f(i-1)-f(i))- 2k_{\text{off}} (f(i)-f(i+1))   \\
    &- k_{-} \left((i-1)f(i) - 2\sum_{j=i+1}^{\infty}f(j)\right) - rf(i).
\end{split}
\label{eq:pi_evolution}
\end{equation}
The balance equation for the steady state is thus
\begin{equation}
    2k_{+} m_0 f(i-1) + 2k_{\text{off}} f(i+1) + 2k_{-}\sum_{j=i+1}^{\infty}f(j)= (2k_{+} m_0 + 2k_{\text{off}} + r + (i-1)k_{-})f(i).
\end{equation}
The steady state distribution here is no longer solved by the simple geometric distribution ansatz, as the term due to fragmentation, $(i-1)k_{-}f(i)$, contains explicit $i$ dependence so no single characteristic equation for the distribution can be obtained.

Instead, we can numerically solve for the steady state, by truncating the system at some large aggregate size and solving the resulting finite set of coupled equations. We consider aggregates of up to size $N$ and write $\mathbf{f} = (f(n_{C}), f(n_{C}+1), \dots, f(N))^{T}$ so that the set of equations described by \eqref{eq:master_eq} now become
\begin{equation}
    \dot{\mathbf{f}} = A\mathbf{f} + \mathbf{q},
    \label{trunc_matrix}
\end{equation}
where
\begin{equation}
    q_i=k_nm^{n_C}\delta_{i,n_C}
\end{equation}
and where $A_{ij}$ is the contribution of aggregates of size $j$ to the rate of change of the population of aggregates of size $i$, which is defined to be
\begin{equation}
    A_{ij} = 2k_+m(\delta_{j,i-1}-\delta_{j,i}) + 2k_{\rm off}(\delta_{j,i+1}-\delta_{j,i}) + k_-\left[2\mathbf 1_{j>i}-(i-1)\delta_{ij}\right] -r\delta_{ij} + k_2m^{n_2}\delta_{i,n_2}\,j
\end{equation}
and these expressions are valid for $i, j \in {n_C, n_C+1, \dots, N}$. Note that here, the matrix indexing starts counting from $n_C$. 

As an explicit example, consider ($n_C=n_2=2$) and truncate the system at $N=4$, such that
\begin{equation}
\mathbf{f}
=
\begin{pmatrix}
f_2\\
f_3\\
f_4
\end{pmatrix}.
\end{equation}
For compactness, define
\begin{equation}
    u=2k_{+}m, \qquad d=2k_{\mathrm{off}}, \qquad g=k_{-}, \qquad b=k_2m^{2}, \qquad a=k_nm^2.
\end{equation}
Equation~\eqref{trunc_matrix} then becomes
\begin{equation}
    \frac{\mathrm d}{\mathrm dt}
\begin{pmatrix}
f_2\\
f_3\\
f_4
\end{pmatrix}
=
\begin{pmatrix}
-u-d-g-r+2b & d+2g+3b &
2g+4b \\[4pt]
u & -u-d-2g-r & d+2g \\[4pt]
0 & u & -u-d-3g-r
\end{pmatrix}
\begin{pmatrix}
f_2\\
f_3\\
f_4
\end{pmatrix}
+
\begin{pmatrix}
a\\
0\\
0
\end{pmatrix}.
\label{eq}
\end{equation}
Elongation from the largest aggregates is treated as loss from this truncated system.

\begin{figure}
    \centering
    \includegraphics[width=0.9\linewidth]{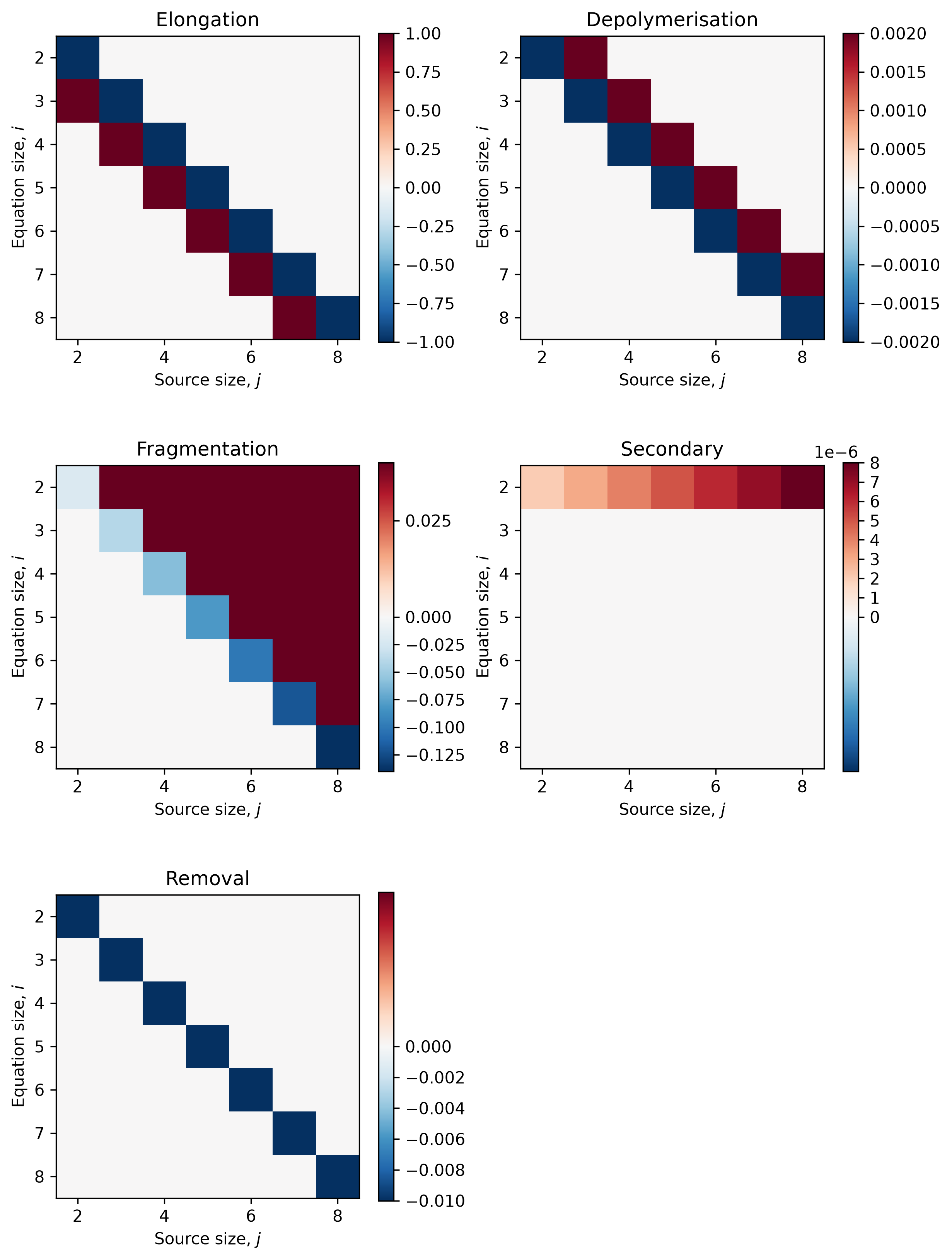}
    \caption{Visualisation of how different processes contribute to the overall rate matrix. The values in the matrix are: $k_n= 10^{-6}$, $k_2= 10^{-6}$, $k_+ = 0.5$, $r=0.01$, $m=1$, $n_c=n_2=2$, $N=1000$, $k_{\mathrm{off}}=0.001$, $k_-=0.02$. }
    \label{fig:rate_matrix}
\end{figure}

\begin{figure}
    \centering
    \includegraphics[width=0.95\linewidth]{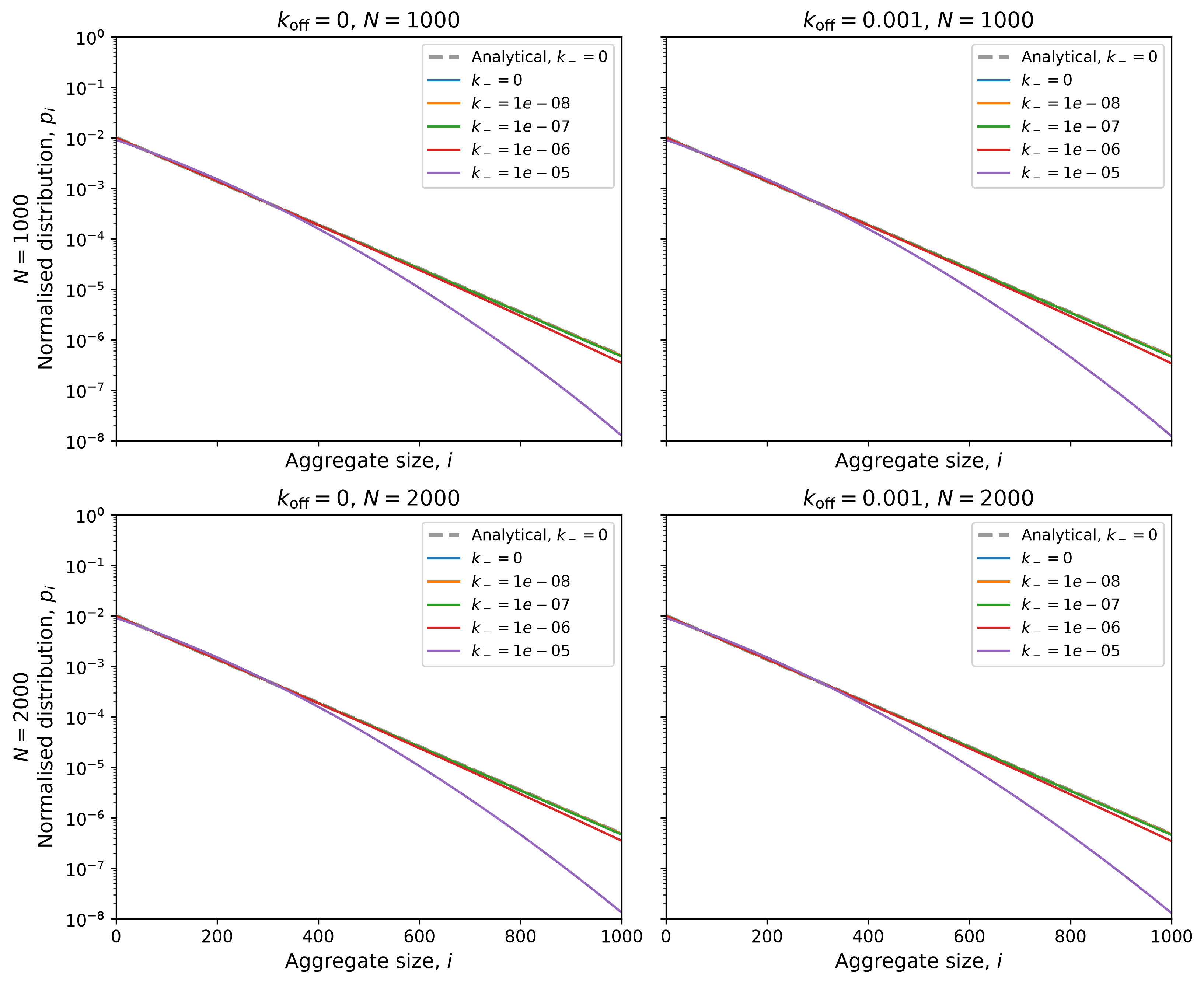}
    \caption{Stationary distributions for truncated systems showing the role of different $k_{-}$ values. When fragmentation becomes significant, the size distributions deviate from a geometric distribution and decay more rapidly with length, giving downwards curvature on the log-linear plot.  The bottom row shows that this is not an artifact of truncation. Parameters used: $k_n= 10^{-6}$, $k_2= 10^{-6}$, $k_+ = 0.5$, $r=0.01$, $m=1$, $n_c=n_2=2$, $N=1000$ (top row), $N=2000$ (bottom row).
}
\label{fig:stat_dists}
\end{figure}

Figure~\ref{fig:rate_matrix} graphically illustrates the contribution of the different processes to this matrix.
We can solve the linear system for a steady state to give
\begin{equation}
    \mathbf{f}_{s} = -A^{-1}\mathbf{q}
\end{equation}
which we can use to plot the steady state aggregate size distribution as a function of the different rate parameters. This is shown in Figure~\ref{fig:stat_dists} for different depolymerisation and fragmentation rates. Crucially, we note that the fragmentation causes the concave curvature that reduces the concentration of large aggregates away from the linear slope of a geometric distribution on the log-linear plot. The presence of such curvature can thus be used as an identifier of the presence of fragmentation.

\section{Application in living system}

\subsection{Length distributions as a Proxy for Cell State}
\label{sec:linking_mass_and_length}
As discussed above and in Cotton et al.~\cite{Cotton2026JCP}, by considering the steady state aggregate concentration in a cell, one obtains two steady states: a ``healthy'' state where aggregates are being removed effectively and are present at low concentrations, and a ``diseased'' state where removal processes are saturated and aggregate concentrations are high. This saturation of removal alone gives rise to differences in the length distribution from healthy and diseased cells. In practice, healthy and diseased cells may also differ in other aspects, for example diseased cells may suffer from impaired protein quality control machinery due to damage caused by the toxic protein aggregates. These secondary feedback effects will tend to further exacerbate differences between healthy and diseased states. As these effects alter the relative balance between the rates governing the size distributions, measurements of aggregate sizes in a biological sample can, ideally, be used to infer the presence of different cells states and the relative rates of the dominant processes.

\begin{figure}
    \centering
    \includegraphics[width=0.9\linewidth]{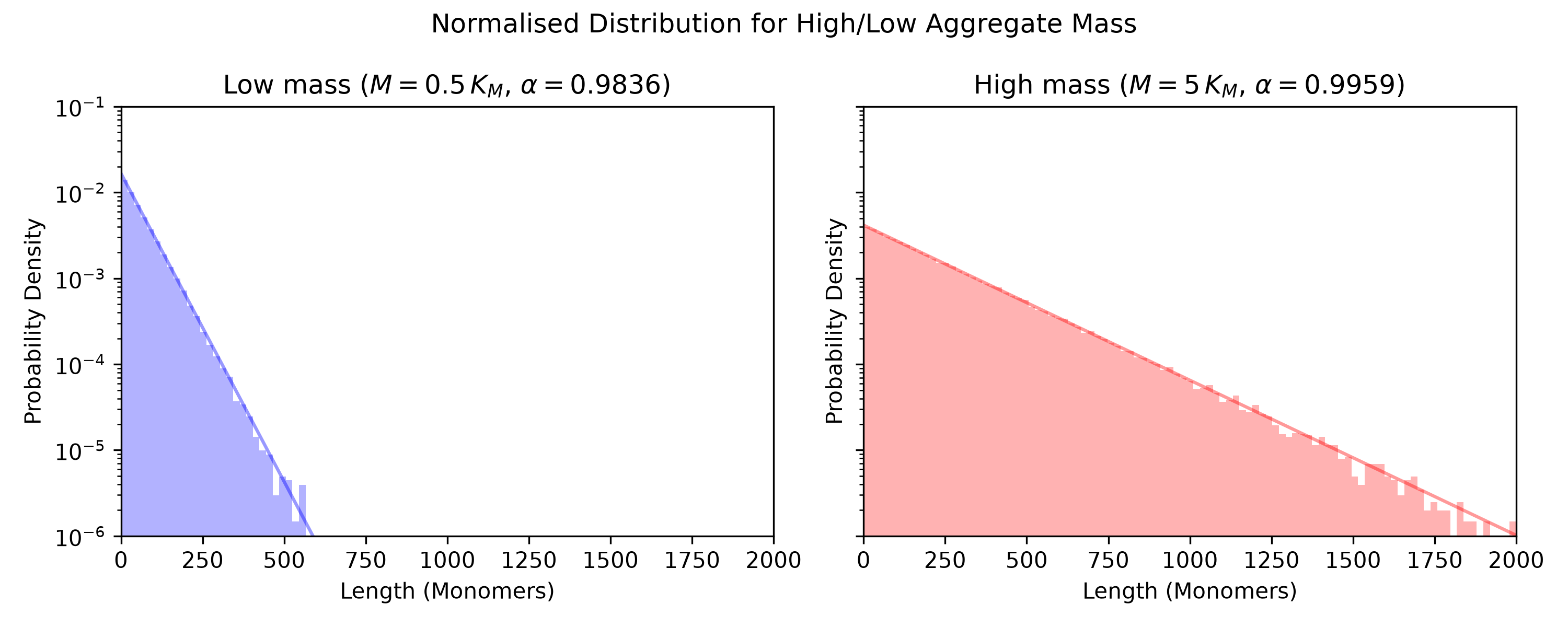}
    \caption{Normalised aggregate length distributions at low and high aggregate mass. Histograms show the normalised distribution the corresponding analytic elongation and removal only geometric distribution. (Left, blue) Low aggregate mass, $M = 0.5 K_M$, giving $\alpha = 0.9836$ and a mean length of around $60$ monomers. (Right, red) High aggregate mass, $M = 5 K_M$, giving $\alpha = 0.9959$ and a mean length of around $240$ monomers. Parameters: $k_+m_0 = 0.1$, $K_M = 2 \times 10^{-6}$, $\lambda = 5 \times 10^{-3}$.}
    \label{fig:twostate}
\end{figure}

To illustrate how aggregate sizes can report on low and high aggregate cell states, consider how the shape of the geometric length distribution links the decay parameter $\alpha$ to the aggregate mass within a cell.  This is due to the functional form of the removal term, $r$, which saturates in the total aggregate mass as given in equation \eqref{def_r}. Thus in the elongation and removal only model, we find
\begin{equation}
    \alpha = \frac{2 k_{+} m_0 (K_{M}+M)}{\lambda  K_{M}+2 k_{+}m_0 (K_{M}+M)}
\end{equation}
and so there is a one-to-one mapping between the geometric decay rate and positive values of aggregate mass. To illustrate this point we can simulate the aggregate length distributions from two different aggregate masses and the corresponding distributions reveal the different aggregates masses even with all other system parameters constant, see Figure \ref{fig:twostate}.

\subsection{Aggregate Size Distributions from Two Distinct Cell States}


In many experimental modalities, tissue samples are homogenised before the length distributions can be measured and as such, we would expect the aggregate size distribution from the homogenate to be made up of aggregates from a range of healthy and pathological cells. Our model as outlined above, along with evidence from histopathology, suggests that cells may in fact group largely into only two distinct states, healthy and diseased, separated by a sudden tipping point. We therefore consider here a system where the length distribution arose from two distinct states which we refer to as healthy and pathological. However, the conclusions generalise to situations where there are more than two distinct states.

\begin{figure} [h]
    \centering
    \includegraphics[width=\linewidth]{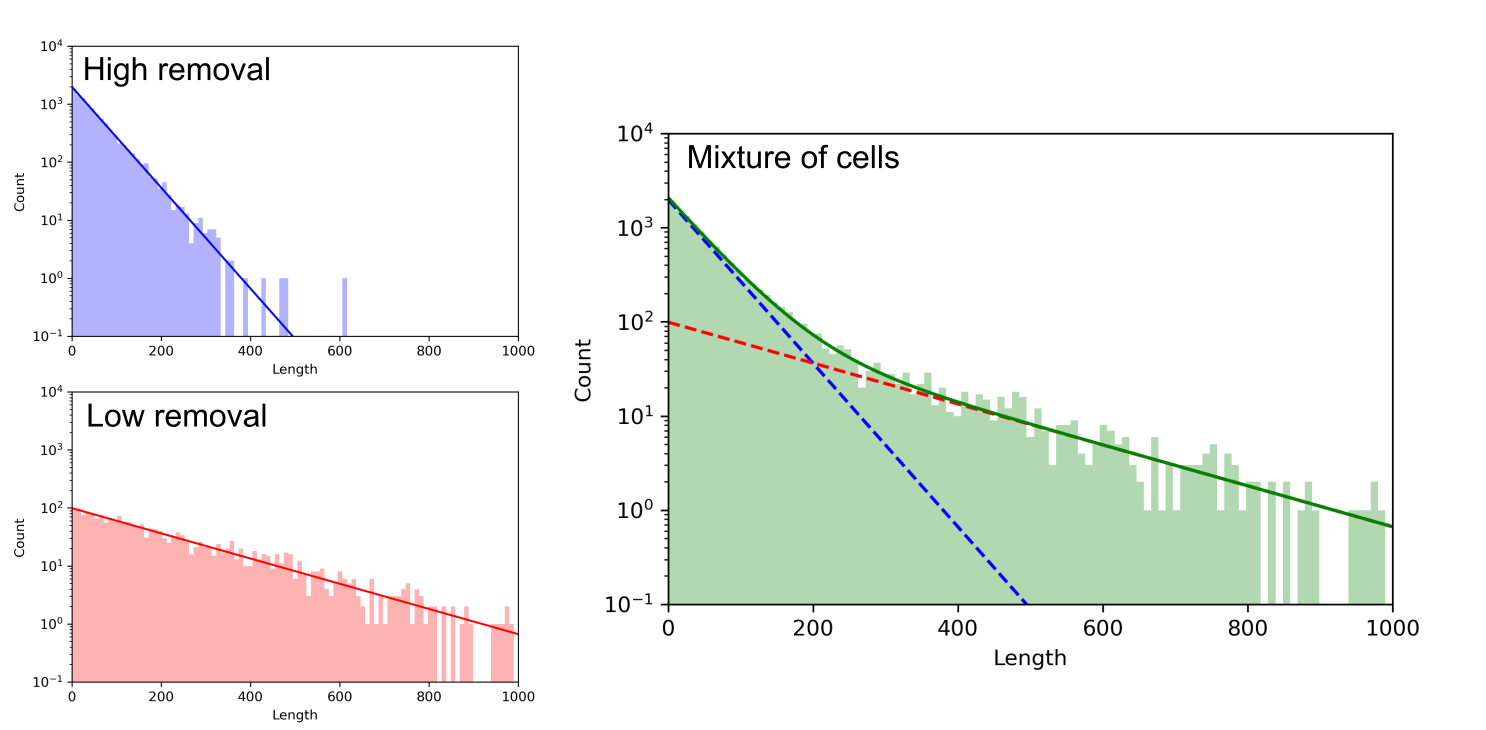}
    \caption{Left: Artificially generated length distribution for a high removal rate ($\alpha = 0.98$), a low removal rate ($\alpha = 0.995$). Right: a mixture of the length distributions from high and low removal rate environments, with 83\% of aggregates from the low removal environment ($\beta=0.83$). The solid green line is equ.~\ref{eq:twoPop}.  Note the curvature and dominance of the low removal rate aggregates at larger lengths.}
    \label{fig:example_distributions_twocells}
\end{figure}

For a mixture of two cellular states, the probability of observing an aggregate of size $i$ is
\begin{equation}
    p_i = (1-\beta)p_i^{h} + \beta p_i^{d}
    \label{eq:twoPop}
\end{equation}
where $p_i^{h}$ ($p_i^{d}$) is the normalised aggregate size distribution in healthy (pathological) cells, which is simply given by the normalised geometric distribution using the decay factor, $\alpha_{H}$ ($\alpha_{D}$). The variable $\beta$ is the fraction of aggregates in the tissue that are from cells in the pathological state. Plotting the resulting combined distribution on a log-linear plot again, the contributions from the two cell states can easily be identified, see Figure~\ref{fig:example_distributions_twocells}. The diseased population gives rise to a long tail, whereas the healthy population dominates at the smaller aggregate sizes. This effect on distinct parts of the distribution makes for easy qualitative assessment of the contribution of the two states, and also allows us to extract the three parameters determining the system, $\alpha_H$, $\alpha_D$ and $\beta$, from experimental data.

In practice, the absolute values of the relative removal rate $\tilde{r}$ may not be easy to interpret, however, using the two different decay factors we can compare the relative removal in health and disease. We define the ratio of the relative removal between health and disease, $r_C$, as
\begin{equation}
    r_C = \frac{\tilde{r}_{d}}{\tilde{r}_{h}} = \frac{\alpha_{D}^{-1}-1}{\alpha_{H}^{-1}-1}.
    \label{eq:relrem}
\end{equation}
This quantity can for example be used to determine by how much the removal rate has decreased between the healthy and diseased states. For example, in the simple case of negligible fragmentation and depolymerisation, using equation~\eqref{eq:alphadef} we obtain
\begin{equation}
    r_C =\frac{r_D }{r_H}\cdot\frac{ k_{+,H}m_{0,H}}{ k_{+,D}m_{0,D}}.
    \label{eq:relrem_simple}
\end{equation}



\subsection{Robustness in practical applications}
To judge how useful a determination of $r_C$ will be in practice, we need to investigate how robust it is to uncertainties in the measured length distributions and how sensitive it is to the assumptions made about the mechanisms.

\subsubsection{Sensitivity to accurate absolute length measurement}
All our modelling above considers aggregate size in terms of the number of monomeric species in the aggregate. By contrast, most measurement techniques will yield physical sizes in nm, so a conversion between physical size and monomer number is usually required. Depending on the measurement technique, assumptions can be made about how to convert between the two quantities, such as using the known beta-sheet separation of 0.5 nm in an amyloid fibril~\cite{Fitzpatrick2017}. However, such conversions tend to be inaccurate, so any derived mechanistic parameter needs to be insensitive to the exact size-to-number conversion factor.
We now demonstrate that the relative removal rates extracted are indeed robust with respect to the precise conversion factor used to relate physical size to monomer number.

Suppose we measure the relative removal rate in two systems, $A$ and 
$B$, or systems in two different states e.g. health and disease. We can compare these rates with the following ratio:
\begin{equation}
    r_C = \frac{\tilde{r}_{A}}{\tilde{r}_{B}} = \frac{\alpha_{A}^{-1}-1}{\alpha_{B}^{-1}-1}.
    \label{eq:clearRat}
\end{equation}

As aggregates tend to be hundreds of monomers in size, $\alpha$ tends to be close to 1, so we define $\alpha = 1 - \epsilon \delta$, with $\delta \sim \mathcal{O}(1)$. Substituting this into equation \eqref{eq:clearRat}, we find
\begin{equation}
    \frac{\tilde{r}_{A}}{\tilde{r}_{B}} = \frac{\alpha_A^{-1} - 1}{\alpha_B^{-1} - 1} = \frac{\delta_A}{\delta_B}.
    \label{eq:exact}
\end{equation}

Now consider the effect of an incorrect conversion from physical size to monomer number, represented by a constant scaling factor $q$. Under this error, the decay parameter is misestimated as $\tilde{\alpha} = \alpha^{q}$ leading to an incorrect estimate $r_C$:
\begin{equation}
    \frac{\tilde{r}_A}{\tilde{r}_B} = \frac{\tilde{\alpha}_A^{-1} - 1}{\tilde{\alpha}_B^{-1} - 1} = \frac{\alpha_A^{-q} - 1}{\alpha_B^{-q} - 1}.
\end{equation}

As aggregates are typically $\sim100$s of monomers long, we expect $\alpha\approx1$ and $\epsilon \ll 1$. Expanding the misestimated ratio in $\epsilon$, we find
\begin{equation}
    \frac{\alpha_A^{-q} - 1}{\alpha_B^{-q} - 1} 
    = \frac{(1 - \epsilon \delta_A)^{-q} - 1}{(1 - \epsilon \delta_B)^{-q} - 1} 
    = \frac{\delta_A}{\delta_B} + \frac{(q - 1)\delta_A(\delta_A - \delta_B)}{2\delta_B} \epsilon + \mathcal{O}(\epsilon^2).
\end{equation}

To leading order in $\epsilon$, this expression reproduces the exact ratio in Eq.~\eqref{eq:exact}. Therefore, the measurement of $r_C$ is robust to errors in the absolute conversion from physical size to monomer count, as long as the aggregates are large.


\subsubsection{Detection of change in removal rate in the presence of depolymerisation}
Having established the robustness with respect to accuracy of the size measurements, we now consider how the presence of depolymerisation or fragmentation may alter our interpretation of $\alpha$ as a measure for the rate of removal relative of the rate of aggregate growth. First, the presence of a fragmentation process affects the shape of the size distribution, leading to downwards curvature. This feature is easy to identify, however, fragmentation seems to be rare and in our analysis of size distributions from human samples we have yet to encounter such a sample.

\begin{figure}
    \centering
    \includegraphics[width=0.7\linewidth]{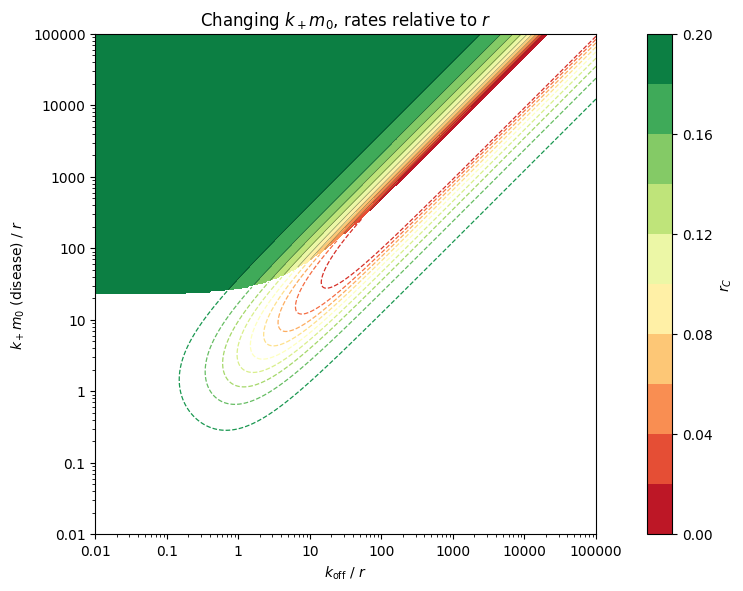}
    \caption{Plot of $r_C$ as given by equation~\eqref{eq:relrem}  and $\alpha$ as given by equation~\eqref{eq:alpha_depol} when $k_+m_0(\mathrm{health}) = 0.2 k_+m_0(\mathrm{disease})$ and the other rates are the same in the diseased and healthy states. The desired value for $r_C$ that correctly reports on the change in the aggregate growth rate,  $k_+m_0$, is thus 0.2, shown as green in the contour plot. The biologically relevant regions in which $\alpha_D>0.9$ and $\alpha_H>0.9$ is shown in solid colours, the contours outside this region are shown dashed and not filled.}
    \label{fig:koff_effect_lambda_unchanged}
\end{figure}

By contrast, in the presence of a depolymerisation process, the size distribution remains a geometric distribution, so this case cannot be easily identified and we thus need to consider how it may lead to misinterpretations of $\alpha$ and $r_C$.
To investigate this, we compare the general expression for $r_C$ in terms of the $\alpha_H$ and $\alpha_D$, equation~\eqref{eq:relrem}, with the ratio of removal and formation rates (which is equivalent to $r_C$ in the case of negligible depolymerisation as in equation~\eqref{eq:relrem_simple}). The latter is the quantity we will be interested in for most experimental contexts, so we want to determine how well it is approximated by $r_C$ even when there is significant depolymerisation. To remain in a meaningful regime of aggregate sizes, we require $\alpha>0.9$ in both the healthy and diseased populations and then consider two limiting cases.

First, assume that the difference between health and disease is due to a difference in the growth rate $k_+m_0$ (i.e. either the rate constant or the monomer concentration is altered), with $k_+m_0(\mathrm{health}) = 0.2 k_+m_0(\mathrm{disease})$. The other rates are assumed to be the same in the two states. We then investigate the behaviour of this system as the importance of the different rates is altered, Figure~\ref{fig:koff_effect_lambda_unchanged}. We normalise all the rates to the removal rate to reduce the degrees of freedom to two and enable us to show the behaviour in a single contour plot. For the majority of parameters, $r_C$ correctly reports on the change in aggregate formation rate between health and disease. Only when both depolymerisation and growth are large, i.e. when removal is negligible, does $r_C$ deviate and over-estimate the difference between healthy and diseased states.

\begin{figure}
    \centering
    \includegraphics[width=0.7\linewidth]{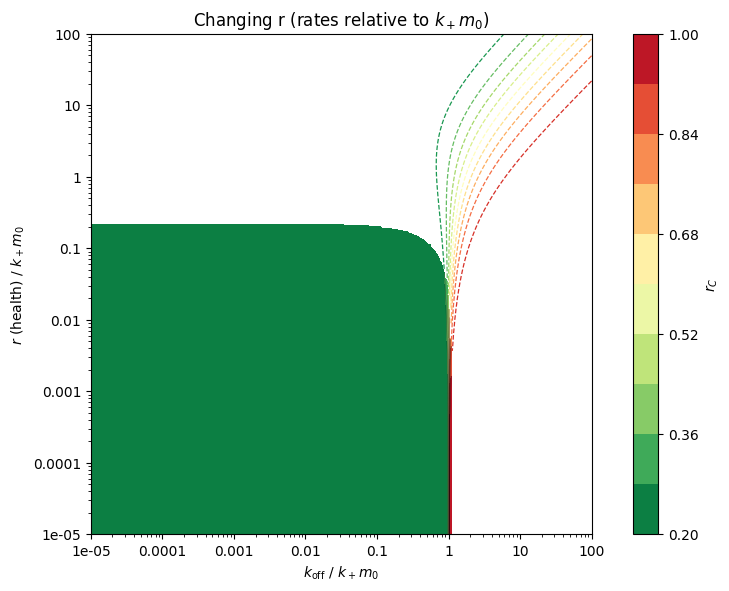}
        \caption{Plot of $r_C$ as given by equ.~\eqref{eq:relrem}  and $\alpha$ as given by equ.~\eqref{eq:alpha_depol} when $r(\mathrm{disease}) = 0.2 r(\mathrm{health})$ and the other rates are the same in the diseased and healthy states. The desired value for $r_C$ that correctly reports on the change in the aggregate removal rate,  $r$, is thus 0.2, shown as green in the contour plot. The biologically relevant regions in which $\alpha_D>0.9$ and $\alpha_H>0.9$ is shown in solid colours, the contours outside this region are shown dashed and not filled. }
    \label{fig:koff_effect_kon_unchanged}
\end{figure}

Second, consider the case when the difference between health and disease is due to a difference in the removal rate constant $r$, with $ r(\mathrm{disease}) = 0.2 r(\mathrm{health})$.  The other rates are the same in the two states. We normalise all the rates to the removal rate to reduce the degrees of freedom to two and enable us to show the behaviour in a single contour plot, Figure~\ref{fig:koff_effect_kon_unchanged}. Again, we find that $r_C$ is an accurate reporter of the changes between health and disease unless the rates of aggregate growth and depolymerisation are essentially the same and removal is negligible. 

In summary, $r_C$ is a robust reporter of the changes in relative removal between two states as long as some removal of aggregates is taking place.

\section{Conclusions}
In conclusion, we have shown that in the presence of a removal process, the large aggregate size distribution is geometric in the absence of a fragmentation process. The decay rate of this geometric distribution is determined by the balance of aggregate growth and removal and measurements of the size distribution can thus give mechanistic insights into the balance of these two crucial processes. We furthermore show that the size distributions from samples that contain different cell types or states, as will often be the case in practice, can provide information on the proportions and relative rates in the different subpopulations. We derive the $r_C$, the ratio of relative removal rates as a key quantity to compare systems and show that it is robust to experimental inaccuracies and slight variations in the aggregation mechanism. This approach to analyse of length distributions provides a way to extract quantitative mechanistic information from human samples through super-resolution microscopy.


\bibliographystyle{unsrt}
\bibliography{references_used}

\end{document}